\documentclass[letterpaper, comsoc]{IEEEtran}

\usepackage{pdfpages}
\usepackage{graphicx}
\usepackage{epsf} 
\usepackage{subfig}
\usepackage{algorithm}
\usepackage{fancyhdr}
\usepackage[english]{babel}
\usepackage{booktabs}
\usepackage{array}

\ifCLASSOPTIONcompsoc
  \usepackage[nocompress]{cite}
\else
  \usepackage{cite}
\fi

\usepackage{hyperref}
\hypersetup{draft}

\newboolean{showcomments}
\setboolean{showcomments}{false}
\ifthenelse{\boolean{showcomments}}
{ \newcommand{\mynote}[3]{
    \fbox{\bfseries\sffamily\scriptsize#1}
    {\small$\blacktriangleright$\textsf{\emph{\color{#3}{#2}}}$\blacktriangleleft$}}}
{ \newcommand{\mynote}[3]{}}

\newcommand{\jpb}[1]{\mynote{jb}{#1}{green}}

\AtBeginDocument{%
  \providecommand\BibTeX{{%
    \normalfont B\kern-0.5em{\scshape i\kern-0.25em b}\kern-0.8em\TeX}}}

\IEEEoverridecommandlockouts
\begin{document}

\title{Impact of Geo-distribution and Mining Pools on Blockchains: A Study of Ethereum\\\small{Practical Experience Report}}


\author{\IEEEauthorblockN{Paulo Silva\IEEEauthorrefmark{1},
David Vav\v{r}i\v{c}ka\IEEEauthorrefmark{1}, Jo\~ao Barreto\IEEEauthorrefmark{1} and
Miguel Matos\IEEEauthorrefmark{1}}\\
\IEEEauthorblockA{INESC-ID \& IST. U. Lisboa\\
\IEEEauthorrefmark{1}
\{paulo.mendes.da.silva, david.vavricka, joao.barreto, miguel.marques.matos\}@tecnico.ulisboa.pt}}

\onecolumn

© 2020 IEEE.  Personal use of this material is permitted.  Permission from IEEE must be obtained for all other uses, in any current or future media, including reprinting/republishing this material for advertising or promotional purposes, creating new collective works, for resale or redistribution to servers or lists, or reuse of any copyrighted component of this work in other works.\\

If you'd like to cite this work, please use the reference below:\\

\noindent\emph{\textbf{Impact of Geo-distribution and Mining Pools on Blockchains: A Study of Ethereum.\\
Paulo Silva,
David Vav\v{r}i\v{c}ka, Jo\~ao Barreto and
Miguel Matos.\\
50th IEEE/IFIP International Conference on Dependable Systems and Networks (DSN), 2020.}}

\twocolumn

\newpage

\pagestyle{empty}
\maketitle
\thispagestyle{empty}

\begin{abstract}
Given the large adoption and economical 
impact of permissionless blockchains, the complexity 
of the underlying systems and the adversarial environment in which they operate, 
it is fundamental to properly study and understand the emergent behavior 
and properties of these systems. 
We describe our experience on a detailed, one-month study of 
the Ethereum network from several geographically 
dispersed observation points. We leverage multiple geographic 
vantage points to assess the key pillars of Ethereum, 
namely geographical dispersion, network efficiency, 
blockchain efficiency and security, and the impact of 
mining pools. Among other new findings, we identify 
previously undocumented forms of selfish behavior and 
show that the prevalence of powerful mining pools 
exacerbates the geographical impact on block propagation delays. 
Furthermore, we provide a set of open measurement 
and processing tools, as well as the data set of the 
collected measurements, in order to promote further 
research on understanding permissionless blockchains. 

\end{abstract}


\section{Introduction}


Over the recent years, permissionless blockchains have enjoyed rapid growth and gathered remarkable interest. 
 Permissionless blockchains enable cryptocurrencies and other distributed applications based on smart contracts that promise to
 revolutionize the current payment methods with the
 ambition to eliminate the need for banks or other centralized entities acting as trusted mediators of financial operations. 

At the core of most permissionless blockchains lies a conceptually simple protocol, such as the Proof-of-Work (PoW) consensus proposed by Nakamoto~\cite{Bitcoin}, which specifies the rules on how the blockchain 
should grow and converge.
Despite this simple core mechanism, real-world permissionless blockchain deployments 
such as Bitcoin~\cite{Bitcoin} or Ethereum~\cite{Yellow},
are extremely complex systems composed of several modules and protocols whose properties are intrinsically hard to understand.
To further complicate matters, not all implementations follow the specifications, while different clients have
different default parameters \cite{Parity2019, Geth17DefaultConfig}. 

Additionally, permissionless blockchains have long evolved from a plain organization composed of individual miners, to an ecosystem dominated by mining pools. 
At the time of writing, the top four mining pools have around 60\% and 70\% of Bitcoin's and Ethereum's total network capacity --- the top permissionless blockchains ---  respectively~\cite{btcPool,ethPool}.
Finally, they rely on a large code-base that includes different implementations (and versions) of the software that blockchain nodes and clients run.
It is fundamental to see how they behave, what works, what doesn't work and opportunities for improvement. 

Recent work has performed measurement studies of popular
permissionless blockchains~\cite{InfPropagaion--02, OnAvailability--01, MeasEthPeers, Sirer2018}. Together, these studies have contributed
to a better understanding of how these intricate large-scale systems perform in practice, 
unveiling some findings that were not originally anticipated~\cite{InfPropagaion--02, OnAvailability--01, MeasEthPeers, Sirer2018}.
Still, the general picture attained by these studies is inaccurate and incomplete. This is mostly due to some key limitations: 
(i) relying on a single observation point, therefore neglecting how the geographical distribution of the network affects such measurements~\cite{InfPropagaion--02, OnAvailability--01};
(ii) not considering the recent predominance of mining pools as first-class components in today's blockchain landscape~\cite{InfPropagaion--02, OnAvailability--01, MeasEthPeers};
(iii) ignoring transaction commit time and how it can be negatively impacted by network delays, out-of-order transactions and empty blocks \cite{Sirer2018,MeasEthPeers}.

In this paper, we describe our experience and lessons
learned on implementing and deploying a measurement infrastructure across different
continents, addressing some of the limitations of previous studies.
We focus on the Ethereum blockchain, the second most valuable cryptocurrency. 


%

We implemented and deployed a measurement infrastructure
consisting of several especially modified Ethereum nodes placed across different
continents.
Our measurement nodes run an instrumented variant of the Geth open-source
implementation of Ethereum clients. Each measurement node can connect to the main network and collect the desired measurements from the transactions and blocks it observes.
We analyzed long-running measurements acquired by four measurement nodes which were
deployed in North America (NA), Eastern Asia (EA), Western Europe (WE) and Central Europe (CE). 


We highlight the following key results of our study:

\begin{itemize}

\item We identify the generalized and consistent practice of different forms of selfish behavior that harm the throughput of the main blockchain. To the best of our knowledge, these practices were not documented and/or not empirically studied systematically before.

\item We confirm that the geographical location has a relevant impact on block reception times and that the prevalence of powerful mining pools exacerbates such effect.

\item We provide empirical evidence that the standard 12-block confirmation
rule of Ethereum may not provide the strong probabilistic
guarantees on block finality that are usually assumed in literature.

\item We confirm that some of the metrics collected by previous works still present the same values and report relevant changes to other such metrics.
\end{itemize}

Our main contributions are the following:

\begin{itemize}
\item A one-month study of the Ethereum network from several geographically dispersed observation points
\item A set of open-source measurement and processing tools that allows other researchers to reproduce our observations and/or perform similar studies in other blockchain systems
\item The collected data set which might enable other researchers to do 
other findings in our collected data or refine 
our observations\footnote{Tools and data set available at: https://angainor.science/ethmeasure}
\end{itemize}


The remainder of this paper is organized as follows.
\S\ref{methodology} describes our measurement infrastructure and methodology.
\S\ref{Analysis} presents the main results of our study, while
\S\ref{relwork} provides essential related work.
\S\ref{Lessons} describes the lessons learned from our results
and proposes lines for mitigating the major threats that our study identified.
Lastly, \S\ref{conclusions} draws our final conclusions.

\section{Methodology}
\label{methodology}

In this section, we describe our methodology.
Recall that our goal is to assess the emerging behavior of Ethereum according to several key aspects, namely: the impact of geographical dispersion, network efficiency, blockchain efficiency, security, and the impact mining pools have on each of these.
To achieve this, we created a modified Ethereum client that collects the metrics of interest and deployed it over three continents.
We used Geth version $1.8.23$ as the basis for our measurements. 
The rationale behind choosing Geth in favor of the other available client implementations is that Geth is Ethereum's reference implementation
and the most widely used client, with more than 74\% of the user share\cite{Ethernodes2019}. The client was instrumented to capture and log all incoming network messages, hence allowing us to collect information about incoming transactions, blocks, and peer connection requests.
Each measurement is logged to a dedicated log file together with a local timestamp.
We collected 600 GB of raw logs and analyzed them using \textit{pandas} \cite{Pandas2019} and \textit{NumPy} for Python \cite{NumPy2019}.

Our modifications to Geth entailed adding and adapting roughly 1,000 lines of code.
We used Geth's default settings  except for the number of peers we can connect to, which we set to unlimited in order to observe as much information from the network as possible.
These settings are identical to the measurement client configuration used in Weber et al.~\cite{OnAvailability--01}, which allows us to compare our results with theirs.
Note that, apart from the instrumentation effort and the number of connected peers, no other changes have been made to Geth --- in particular, our client behaves like any other client in the network and thus it is indistinguishable from any other regular client.
This is fundamental to obtain unbiased results.


The instrumented version of Geth was deployed in computing instances located in North America, Eastern Asia, Western Europe and Central Europe. 
They were connected directly to the Internet backbone with a network throughput of at least 8 GB/s. 
A detailed description of the specification of each machine can be found in Table~\ref{tab:machines}.
These specifications are well above the minimum requirements to run an Ethereum client so we do not expect any biases due to poor hardware performance~\cite{ethBook}. 
Each machine used the Network Time Protocol (NTP) 
for clock synchronization. NTP provides offsets lesser than 100ms in 99\% of cases and lesser than 10ms in 90\% of cases\cite{NTP}.

Part of our study involves analyzing propagation delays
in the Ethereum network, with a particular focus on blocks.
We adapt the method proposed by Decker et al. \cite{InfPropagaion--02}, which exclusively relies on timestamps generated by our measurement nodes to compute the block propagation delay. More precisely, we define the propagation delay of a block as the time difference between the first observation of that block at any instance of a measurement node and the times of arrival on the remaining measurement nodes. 
Note that this is an approximated method since (i) it does not measure the time it takes to propagate a
transaction or block from the miner
to the first measurement node that received that block; and (ii) the accuracy of our measurements is always bounded by the accuracy of NTP. 
We take this limited accuracy into consideration whenever relevant.


\begin{table}
\begin{tabular}{m{1cm} m{3.2cm} m{1cm} m{1cm}}
\hline
Location & CPU & RAM (GB) & Bandwidth (Gbps) \\
\hline
NA  & 4x  Intel~Xeon~2.3~GHz & ~~15    & ~~8   \\
EA  & 4x  Intel~Xeon~2.3~GHz & ~~15    & ~~8   \\
CE  & 4x  Intel~Xeon~2.4~GHz & ~~~8    & 10  \\
WE  & 40x Intel~Xeon~2.2~GHz & 128    & 10  \\ \bottomrule
\end{tabular}
\caption{Specifications of the measurement infrastructure.}
\label{tab:machines}
\end{table}



We performed the measurements from April 1\textsuperscript{st} 2019 to May 2\textsuperscript{nd} 2019, with each machine connected to more than 100 peers at any moment.
Additionally, we did a complementary measurement on the WE instance with the default number of 25 peers. 
This measurement was meant to capture the behavior of an Ethereum client with default settings and took place from May 2\textsuperscript{nd} to May 9\textsuperscript{th}.
\textit{Ethical Considerations:} The machines and Geth client we deployed follow exactly the Ethereum protocol rules and thus have no negative impact on the behavior of the Ethereum network.
The data we collected is publicly available to anyone that connects to the Ethereum network and therefore it does not raise privacy concerns.


\section{Results}
\label{Analysis}

In this section, we present our measurements and discuss the obtained results in face of our expectations and also, whenever appropriate, how they relate to other studies.
During our one-month measurements, we collected data about 216,656 blocks (including forks) with the block numbers ranging from 7,479,573 to 7,680,658.
On top of that, we captured 21,960,051 unique transactions out of which 20,654,578 (94\%) were valid transactions included in main blocks.

Our study focuses on answering a wide set of questions that depend on distinct, yet cross-dependent, facets of the whole Ethereum platform.
We 
follow a bottom-up structure.

 \subsection{Network Efficiency}
\label{subsec:netEff}

The Ethereum network disseminates transactions and blocks using a gossip-based protocol. 
If the network is slow in disseminating transactions this means that end users will observe a large latency in the transactions they submit to the system.
Regarding blocks, a slow block propagation has harsher consequences, since it will lead to more forks as miners are not aware of each other's blocks in time.
Therefore, this section studies the efficiency of Ethereum's network, focusing
on the propagation of blocks.
More precisely, our focus is on understanding how fast the Ethereum network
propagates blocks and on whether it generates significant message redundancy.





\subsubsection{Block and Transaction Propagation Delays}
\label{subsec:block-propag-times}

\begin{figure}[t]
	\centering
	\includegraphics[width=\linewidth]{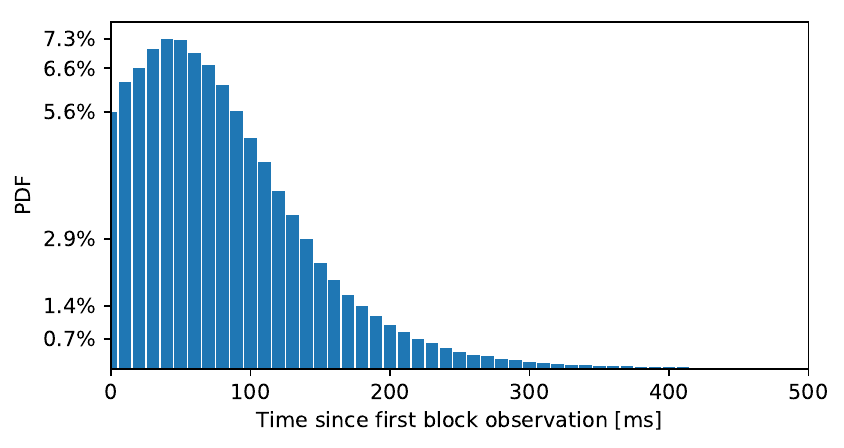} 
	\caption[Block propagation times]{The histogram of times since the first block announcement.}
	\label{fig:information-propagation}
	\vspace{-0.2cm}
\end{figure}

Figure~\ref{fig:information-propagation} depicts the results for block propagation delays.
The median block propagation delay was 74ms and the average was 109ms. 
The propagation delay of the 95\% fastest blocks was under 211ms, and it took 317ms for 99\% of blocks to propagate through the network.
This shows that blocks are propagated several orders of magnitude below the
average inter-block time (currently 13.3s). 

Regarding transaction propagation delays, we did not find them to be affected by geographic location (results not shown due to space constraints).
There are two main factors that explain this observation: i) transactions are small and propagate faster than blocks and within the margin of our measurement error (\S\ref{methodology}), and  ii) transactions tend to be created in a more geographically dispersed fashion (transactions are submitted from a large set of nodes) as opposed to blocks (where just a few miners produce most of them).

\subsubsection{Block reception redundancy}
Blocks are disseminated through two types of messages: either in the form of light announcements (consisting of only the block's hash) or propagated directly (including both header and body).
The dissemination protocol has builtin redundancy mechanisms to tolerate faults and packet loss. 
However, such redundancy comes at the cost of additional network delays and processing overhead.
Therefore, we are interested in knowing how many redundant blocks a node with default settings receives.
Because our measurement nodes are connected to more nodes than the default (\S\ref{methodology}), we performed a subsidiary measurement for this metric, between May 2\textsuperscript{nd} 2019 and May 9\textsuperscript{th} 2019, where an additional measurement node was connected to the default number of peers (25).
\begin{table}[t]
\centering
	\begin{tabular}{ccccccc}
	\toprule
    	{\textbf{Message Type}} & {\textbf{Avg.}} & {\textbf{Med.}} & {\textbf{Top 10\%}} & {\textbf{Top 1\%}} \\ \midrule
    	{Announcements}  & 2.585  & 2 & 5 & 7  \\
    	{Whole Blocks}  & 7.043  & 7 & 10 & 12  \\
    	{Both combined}  & 9.11  & 9 & 12 & 15   \\ \bottomrule
	\end{tabular}
\caption[Redundant Block Receptions]{Redundant block receptions.}
\label{tab:redundantblockreceptions}
	\vspace{-0.2cm}
\end{table}
The results are depicted in Table~\ref{tab:redundantblockreceptions}, which shows that blocks are more often propagated directly rather than via announcements.
The median and mean number of redundant block message receptions is 9.00 and 9.11, respectively, considering both announcements and direct block propagation messages. Even the top~1\% of most redundantly propagated blocks are received just 15 times. 
Eugester et al.~\cite{EpidemicDistr} show that, in networks with failures, it is enough for the gossip protocol to disseminate information to a logarithmic number of neighbors with respect to the total system size.
According to the latest estimation from~\cite{MeasEthPeers}, there are around 15,000 Ethereum peers.
Therefore the measured mean of 9.11 block receptions is close to the optimal value of 10 ($\ln(15,000)$ $\approx$ 9.62).
This is further confirmed by the low propagation times analyzed in \S\ref{subsec:geoImpact}.

 \subsection{Geographical Impact}
\label{subsec:geoImpact}

We now study the impact that geographic location has on block propagation delays. 
This is important because, if some region has lower propagation delays than others, that region has an advantage when mining new blocks, as miners will become aware of the latest blocks faster and thus can start mining the next block ahead of miners in other regions.
\subsubsection{Geographical position influence}
\label{subsec:geo-influence}

The Ethereum network establishes neighboring relationships among peers based on a random node identifier. 
This is independent of the geographic location and therefore, assuming that the network capacity (bandwidth and latency) is evenly distributed among miners, nodes should observe similar propagation delays regardless of their location.
Regarding transaction propagation delays, we did not find evidence that
they were affected by geographic location (results not shown for space limitations). However, we found that block propagation delay is affected by geographic location.

\begin{figure}[t]
 \centering
 \includegraphics[width=\linewidth]{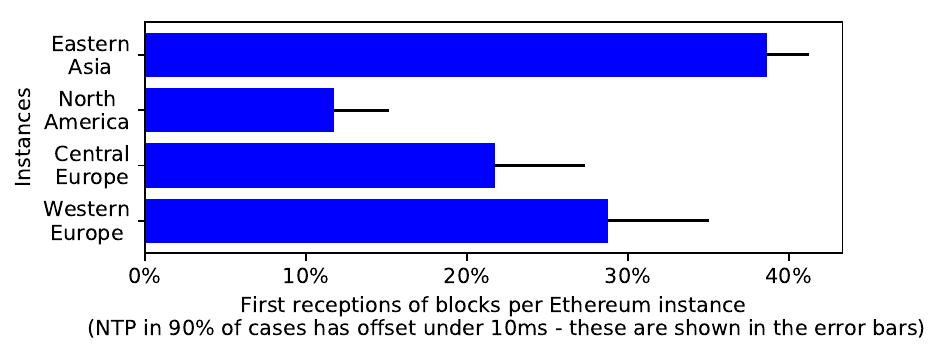}
 \caption[Block propagation]{First new blocks observations with respect to geographical location. The error bars represent the experimental error (\S\ref{methodology}).}
\label{fig:geo-position-vs-block-observ-time}
	\vspace{-0.2cm}
\end{figure}

To assess this, we measured the proportion of times each of our measurement nodes was the first to observe a new block. 
The results are depicted in Figure~\ref{fig:geo-position-vs-block-observ-time}.
The results clearly show that nodes located in EA are the first to receive new blocks most of times ($\approx$40\% of times) whereas nodes in North America are around four times less likely to observe  new blocks first.
Therefore, the geographical location of nodes affects the new block observation times and therefore miners in EA are at an advantage.
The cause of this, as we show in the following measurement, is simply due to the fact that several prominent mining pools operate in Asia and therefore nodes in EA are more likely to receive new blocks first. 


\subsubsection{Mining pools' location}

\begin{figure}[t]
	\centering
	\includegraphics[width=\linewidth]{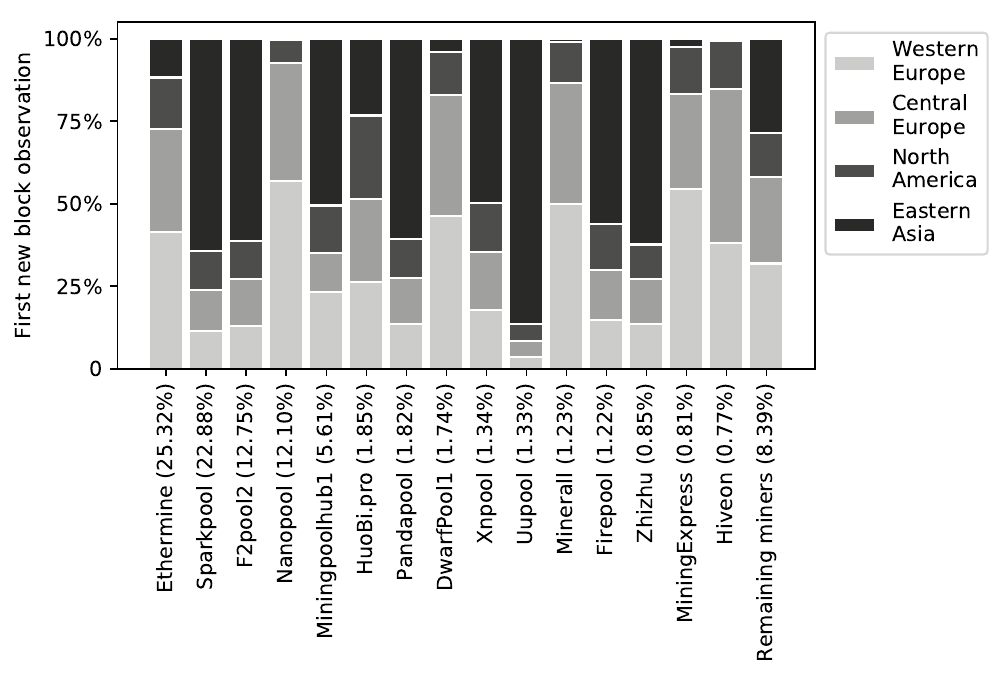} 
	\caption[Block Commit Time]{The influence of a block's origin mining pool on the faster propagation times to various geographical locations. 
	In parentheses, we show the computational power of each mining pool during our experiment.} 
	\label{fig:ptm-4-c}
	\vspace{-0.2cm}
\end{figure}

The emergence of huge mining pools does not only centralize mining power on a few entities, it also centralizes (on a few geographical hot-spots) block propagation. 
To overcome this, and counter the effect observed in the first experiment, mining pools have been known to place gateways in several geographical locations in order to help disseminate their blocks, without disclosing their precise location to avoid attacks \cite{miller2015discovering}. 
To study the impact of this, we measure whether our geographically dispersed measurement nodes capture blocks mined from particular mining pools faster than from others.
The results are depicted in 
Figure~\ref{fig:ptm-4-c}, which shows first new block reception per individual mining pool. 
We consider only the 15 most prominent mining pools, since the fractions of blocks produced by the smaller pools are insignificant. 
The results clearly show that the geographic location of peers affects faster block observation from certain pools, and indicates that the gateways of mining pools are not evenly distributed.
 
\subsection{Blockchain Efficiency}
\label{subsec:blockchainEff}
We now focus on higher level aspects of Ethereum, from the perspective of a blockchain platform.
We focus on aspects that are important to both end users, such as transaction commit time, and miners, such as mining empty blocks.

When an application observes a new block B, it is not safe to immediately consider the state transition given by its transactions, because there is a chance that B might be discarded due to a fork. Thus, applications must wait for a \emph{long enough} suffix of blocks to ensure that the appearance of an alternative heavier chain, not including block B, has a small probability. This property is known as block finality.
Applications choose the probability of chain replacement they are willing to tolerate (i.e., the probability of a block not being final) and wait for enough confirmation blocks to ensure that probability.  
In Ethereum, it is generally accepted that applications should wait for 12 confirmation blocks before considering a block B as final and its transactions as committed~\cite{bez2019scalability, OnAvailability--01, Buterin2015}.


\subsubsection{Transaction commit time}
\begin{figure}[t]
	\centering
	\includegraphics[width=\linewidth]{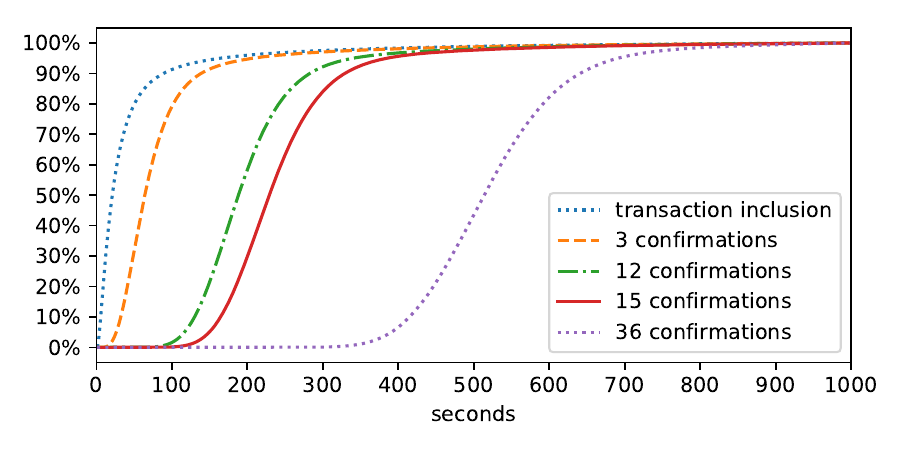}
	\caption[Transaction Commit Time]{Time for transaction inclusion and commit with 3, 12 (default), 15 and 36 block confirmations.}
	\label{fig:tx-commit-time}
	\vspace{-0.2cm}
\end{figure}

We measured the difference between the time when a transaction was first observed by our measurement nodes to the time at which it was included in a block. 
To determine the block confirmation time, 
we also measured how much additional time it took for such a block to be followed by different numbers of blocks in the main chain. 
These are the metrics that mostly affect end users, and they have a direct impact on user perceived latency.
Figure~\ref{fig:tx-commit-time} shows the times of first inclusion of transactions in a block, and the 3-, 12-, 15-, and 36-confirmation block times.
The variants other than the regular 12-blocks case make sense for some applications depending on their requirements regarding block finality probability. 
Besides, as we discuss in \S\ref{subsec:sec}, for applications concerned with the blockchain security, waiting just for 12 blocks might not be enough.
Our measurements revealed that the median waiting time for 12 blocks  was 189 seconds whilst in 2017 it was 200 seconds~\cite{OnAvailability--01}.
The cause for this is that the inter-block time decreased, 
from 14.3~seconds to the current 13.3~seconds~\cite{Etherscan2019}.
This is likely to be related to the Ethereum Constantinople fork that occurred on February 2019 to decrease the inter-block time, which was slowly increasing due to a known hard-coded difficulty bomb \cite{Eips1234, Jameson2019}.

\subsubsection{Transaction reordering}
\begin{figure}[t]
	\centering
	\includegraphics[width=\linewidth]{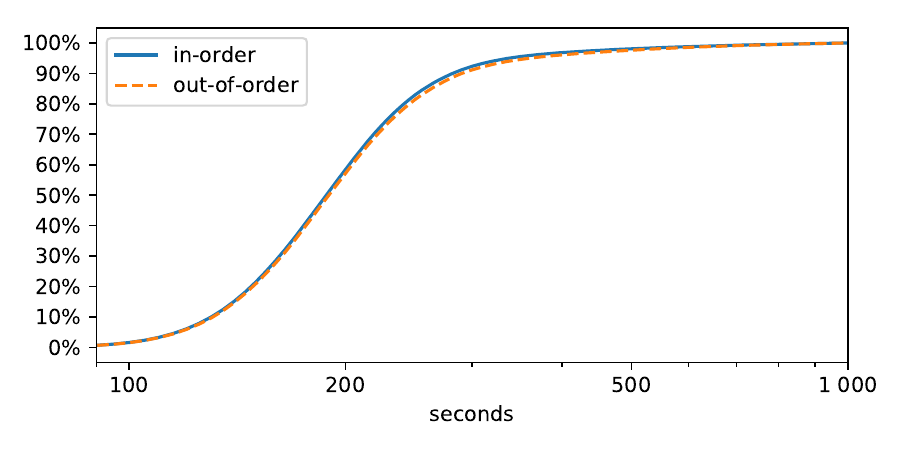}
	\caption[Transaction commit delay based on ordering]{Commit delay (sec) for transaction based on ordering.}
	\label{fig:tx-reord}
	\vspace{-0.2cm}
\end{figure}

The transaction creator stamps every transaction with a monotonically increasing nonce. 
We say that two transactions from the same sender were received out of order when we first observe the transaction with the higher nonce. 
Miners cannot include out-of-order transactions in a block until they receive all foregoing transactions, 
which implies that out-of-order receptions negatively impact transaction commit times, as such transactions must wait for their delayed predecessors before committing.
In 2017, 6.18\% out of all committed transactions were received out-of-order~\cite{OnAvailability--01}.
In our measurements, we observed 11.54\% out-of-order committed transactions, a substantial increase.
We also observed that it takes less than 192 and 325 seconds for 50\% and 90\% of out-of-order transactions to commit.
In comparison, the median time for in-order received transactions is less than 189 seconds and 90\% of these transactions need 292 seconds or less to commit.
The results are depicted in Figure~\ref{fig:tx-reord}.

\subsubsection{Empty blocks}

We now focus on a set of metrics that affect the behavior of the miners.
Blocks have a maximum number of transactions they can include and currently most blocks are at around 80\% capacity~\cite{Etherscan2019}. 
In principle, miners are incentivized to include transactions in a block because they collect the fees associated with each transaction.
However, miners may occasionally decide to create and propagate blocks that
include no transactions. This behavior grants them the possibility of starting to mine earlier than other miners. 
This has interesting consequences.
On the one hand, miners are penalized by not collecting transaction fees.
On the other hand, they still get the mining reward which is, on average, considerably higher.
Besides, empty blocks can be propagated earlier, because miners do not waste time validating transactions, and faster, since they become smaller due to the absence of transactions. 
Overall, these constitute a perverse incentive to mine empty blocks. As a matter of fact, empty blocks are harmful to the network because they increase the commit time of transactions, as transactions that could have been included in an empty block must wait to be included in the next block. 
If a dominant number of miners switched to the selfish strategy of occasionally mining empty blocks, it would be disastrous for the platform.
To assess the impact of this, we measure the number of empty blocks in the network, and the mining pools from which they originate.
The results reveal that 1.45\% are empty blocks (2,921 out of 201,086 total main blocks). This significant fraction of empty blocks decreases transaction throughput, by increasing the transaction commit delay. 

\begin{figure}[t]
	\centering
	\includegraphics[width=\linewidth]{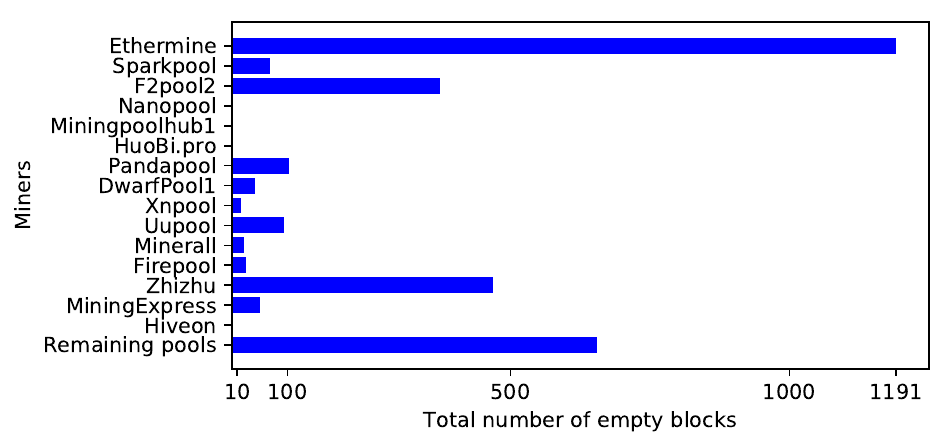} 
	\caption[Block Commit Time]{Empty blocks per mining pool.}
	\label{fig:ptm-3}
	\vspace{-0.6cm}
\end{figure}

Figure~\ref{fig:ptm-3} shows the 15 biggest pools and their share of empty blocks. 
Remarkably, only a small a portion of pools, e.g. Nanopool or Miningpoolhub1, had not mined any empty blocks during our measurement. 
On the other hand, more than 25\% of blocks mined by the Zhizu pool were empty, without a single transaction. 
We also observed a miner whose 6 mined blocks during the experiment were all empty. Etherscan data confirms this miner has systematically only mined empty blocks since its account was created~\cite{Etherscan2019}.

We can therefore conclude that the mining of empty blocks varies substantially across mining pools, which shows that this practice depends on the specific protocols and policies used by each mining pool.
The fact that one major mining pool resorts to this practice frequently might show that the benefits of this selfish behavior are relevant.
This, in turn, may suggest that this behavior may be replicated more aggressively by other mining pools in the future, which will imply higher penalties on the commit delay.

\subsubsection{Blockchain forks}
\label{subsubsec:forks}

In our measurement, we were able to observe nearly all blocks that were created on the network.
This included 485 unrecognized forks that are not reported by popular Ethereum blockchain explorers like Etherscan~\cite{Etherscan2019} or Etherchain~\cite{Etherchain2019}. 




Out of the 216,671 blocks that we captured, 92.81\% of them became part of the main chain, 6.97\% became uncles referenced by some block from the main chain and only 0.22\% of the blocks became unrecognized uncles. 
Table~\ref{tab:forkedblocks} also shows that forks of length one are the most common (97\%) and that the longest forks observed were of length 3. It also shows that forks of length one are very likely to become recognized, i.e. referenced as uncle in some main block. During our measurement, not a single fork longer than 1 became recognized.

\begin{table}[h]   
\centering
	\begin{tabular}{ccccccc}
	\toprule
    	{\textbf{Fork Length}} & {\textbf{Total}} & {\textbf{Recognized}} & {\textbf{Unrecognized}} \\ \midrule
    	{1}  & {15,171}  & 15,100  & 71   \\
    	{2}  & {404}  & 0  & 404   \\
    	{3}  & {10}  & 0  & 10    \\ \bottomrule
	\end{tabular}
\caption[Fork Lengths]{Fork types and lengths.}
\label{tab:forkedblocks}
	\vspace{-0.4cm}
\end{table}

Since 2017, the proportion of forked blocks increased by more than one percent and their lengths increased as well.
Among other possible factors, it is likely that this trend reflects the fact that mean inter-block time (the time between two succeeding blocks) has decreased
by around one second in the last two years~\cite{Etherscan2019}.

\subsubsection{One-miner Forks}
The Ethereum yellow paper defines fork as ``a disagreement between nodes as to which root-to-leaf path down the block tree is the best blockchain''~\cite{Yellow}.
Forks are thus expected to occur when distinct miners disagree on the best blockchain (e.g., on distinct versions of the highest block produced by distinct miners), but not due to a single miner producing distinct blockchains simultaneously (e.g., distinct versions of the highest block).
Surprisingly, we find many instances where a single miner produced several blocks at the same height.
This phenomena, which clearly was not foreseen in the original specification, has a relevant impact today.
In fact, more than 11\% of all forks consisted of a divergence between two blocks from the same miner.

We find that miners produced 1,750 block pairs with a unique block height. They also mined 25 triples of blocks, once mined a 4-tuple and once a 7-tuple of such blocks.
In the case of the 4- and 7-tuples, we believe that these were due to a mining pool partition or another pool malfunction.
In the case of a 3- and 2-tuples, there is a strong reason to suspect of intentional behavior: these forks got recognized as uncle blocks and thus got rewarded in 98\% of the cases.
This phenomenon shows that the uncle block rewarding system, which was intentionally meant to help less powerful miners, is effectively helping the most powerful mining pools to unethically profit from multiple rewards, by mining multiple versions of the highest block in parallel.


 \subsection{Security} 
\label{subsec:sec}

\begin{figure}[h]
	\centering
	\includegraphics[width=\linewidth]{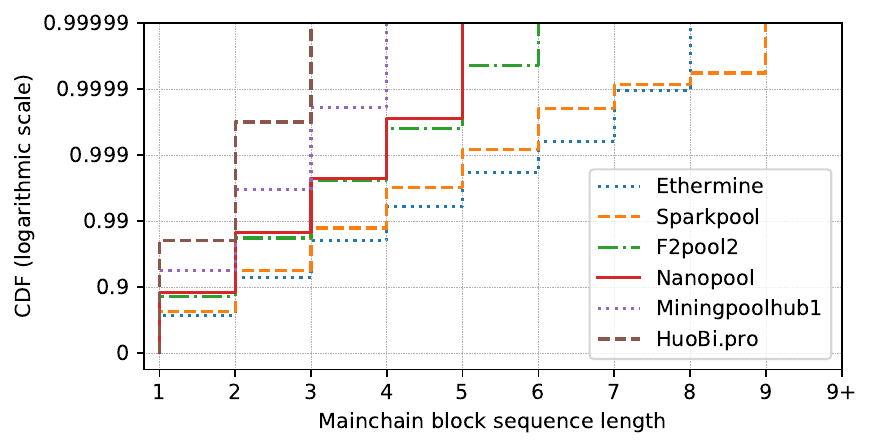}
	\caption[Block Commit Time]{The sequences of consecutive main blocks mined by a unique miner.}
	\label{fig:ptm-1-a}
	\vspace{-0.4cm}
\end{figure}

As previously discussed, mining in Ethereum is performed by a few mining pools that dominate the platform. 
Moreover, miners are free to select which transactions to include in a block and which to discard.
This raises concerns over the security and fairness of the network, as a mining pool might censor transactions from a given user, or perform other attacks such as a double-spend by reverting a suffix of the blockchain.

In Ethereum, a block B is usually considered final when it is followed by a 12 block sequence.
If a mining pool is able to produce more than 12 blocks in row, this means that it can effectively censor the blockchain and perform attacks such as double-spends.
A less severe attack that mining pools can perform is to increase the latency of a given transaction simply by refusing to include it in the sequence of blocks they mine.
This 
enables them to perform a temporary censorship.
To assess the security of the network, from the perspective of block finality, we measured the length of consecutive sequences of blocks created by the same mining pool.
We were interested in the probability of temporary censorship taking place and for how long mining pools would be able to do it.
Our results reveal that Ethereum pools regularly have the opportunity to temporarily censor transactions for more than two minutes, but historically we registered events allowing for 3-minute-long censoring.



During our one-month measurement, we observed that the prominent pools managed to mine sequences of blocks as long as 9. 
The results are shown in Figure~\ref{fig:ptm-1-a} which depicts the lengths of consecutive blocks the top 5 mining pools were able to produce. 
Ethermine 
managed to mine four 8-block long sequences and Sparkpool was able to generate 9-block long sequences twice. 






At the time of our measurements, the accumulated computational power of all Ethermine's miners was 25.9\% of the whole Ethereum platform~\cite{Etherscan2019}. Therefore, the theoretical chance of mining a sequence of 8 consecutive blocks would be $0.259^8 = $ 
$2 \times 10^{-5}$.
During one month, there were 201,086 blocks in the main chain.
With a theoretical chance of $2 \times 10^{-5}$, this means that Ethermine should be able to mine 8 consecutive blocks 4 times per month ($2 \times 10^{-5} \times 201,086 \approx 4$) exactly the value we observed.
In the case of Sparkpool, which has a theoretical chance of mining 9 consecutive blocks of $0.226^9$ it should take at least three months to mine such a sequence ($0.226^9 \times 201,086 \approx 0.3$) however it did so twice in a month.
Since blocks were not announced all together, like in a block withholding attack, and presented an average inter-block time, it is unlikely that Sparkpool performed such an attack~\cite{kwon2017selfish}. It is more likely that the current values that are used to consider a block as final are too optimistic, given the fraction of mining power that is currently held by mining pools.
To further justify this observation, we looked beyond our one-month experiment, and analyzed the whole blockchain.
We observed 102, 41, 4 and 1 sequences of 10, 11, 12 and 14 consecutive blocks, respectively.
The longest sequence ever recorded, consisting of 14 blocks, was mined by Ethermine from block height 5,899,411 to 5,899,424.
We do not know the exact computational power of Ethermine at the time, but if we assume that it was similar to its current power (0.259) the probability of such a long sequence would be around once in 1,000 years.






\section{Related work}
\label{relwork}

A body of work has studied decentralization, the key distinctive feature of blockchain with respect to more traditional centralized approaches, and an important property for high resistance against censorship of individual transactions~\cite{Sirer2018, Luu2017, eyal2018majority, miller2015discovering}. 
Luu \textit{et. al}~\cite{Luu2017} reported that around 80\% of the mining  power in Ethereum resides in less than ten mining pools, which is corroborated by our observations.
Gencer \textit{et. al}~\cite{Sirer2018} showed that both Bitcoin and Ethereum suffer from a centralized mining process, due to mining pools. Miller \textit{et. al}~\cite{miller2015discovering} showed that 75\% of the mining power in Bitcoin resides in just 2\% of the nodes.
Eyal and Sirer~\cite{eyal2018majority} described an attack to Bitcoin preventing decentralization, in which rational miners prefer to join the selfish miners and the resulting colluding group becomes a majority.
Our observations confirm these hypotheses and show that Ethereum mining pools have the power to temporarily censor transactions and harm their commit time by mining long sequences of blocks.

Previous research has focused on the time it takes for a transaction to commit~\cite{OnAvailability--01, Bitcoin, Buterin2015}. 
Nakamoto~\cite{Bitcoin} showed the probability of not replacing a Bitcoin block $B$ containing transaction $t$ with a malicious block $B'$ without $t$ can be made arbitrarily high, whereas Buterin \cite{Buterin2015} studied the corresponding probability for Ethereum. 
In Bitcoin, the probability that is deemed \emph{safe} is achieved after 6 blocks, corresponding to one hour, whereas in Ethereum this is achieved after 12 blocks, corresponding to around 3 minutes~\cite{OnAvailability--01}.
In our one-month observations we observed two instances where a single mining pool was able to mine 9 consecutive blocks twice, and we also observed that over all the blocks ever mined, a mining pool was able to mine a sequence of 14 consecutive blocks.
This means that mining pools are indeed able to censor transactions and rewrite the blockchain, and therefore should raise concerns about the security of the network. 

Previous work attempted to describe the causes of mining empty blocks~\cite{mccorry2018smart, kovst2018transition}. 
In our work, we observed that 1.43\% of Ethereum blocks are empty and most mining pools mined empty blocks, which suggests mining empty blocks pays off.

\section{Lessons learned}
\label{Lessons}
Our experiment shed light on strengths and challenges of the Ethereum network, while offering some surprises.
The low propagation delay we observed can be considered a strength of the Ethereum network.
In contrast, mining pool centralization can lead to challenging and surprising selfish behaviors. 
Commit delays in Ethereum have been improving relatively to the delays reported in prior studies, which can be mostly explained by the adoption of
shorter inter-block times in Ethereum \cite{Etherscan2019}.
Still, we identified selfish behaviors that may place real threats to the
throughput of the system -- most notably, empty blocks and one-miner forks.
They all represent distortions of the incentive model of Ethereum, which
encourage selfish nodes to waste system resources (namely, mining power
and network capacity) in
intentional efforts that do not contribute to the progress of the main
blockchain.
To the best of our knowledge, these behaviors were not anticipated in the
original design of the system \cite{Yellow}. Hence, current implementations
tolerate them.
While our study found relatively scarce occurrences of such
selfish behaviors, with a low impact on the overall throughput of the system,
they were observed consistently over the experiments.
This suggests that these behaviors are profitable to selfish nodes,
hence there is a risk that the frequency and impact of such situations
grows in the future.

Regarding one-miner forks, we argue that the Ethereum protocol should forbid referencing uncles mined by miners that have already mined a main block of the same height. This would -- as our results show -- save around 1\% of the platform's overall computational resources which are currently spent on mining forks, while at the same time giving a higher chance that small miners collect those rewards. The uncle block rate of a mining pool would be effectively slowed down, even if the mining pool tried to use distinct coin addresses for claiming the rewards, since mining power would be split among those addresses. Additionally, we have observed that, in 56\% of cases, mining pools appeared to be using their full mining power for mining distinct versions of the same block (i.e. with the same transaction set) with the same height. In the remaining 44\% of the cases, they were mining different blocks (i.e. with distinct transaction sets) with the same height. This means our solution would effectively deter mining pools from using their full mining power to mine distinct versions of the same block, in more than half of the one-miner fork cases.
Further, producing such blocks could be considered a protocol violation and a miner could be punished by having funds removed from his coinbase account~\cite{xiao2019survey}.
A robust solution to put an end to empty blocks should be designed,
to prevent them from harming the system throughput.
However, this is a challenging endeavor that is left for future work.

Our study also highlights that the emergence of mining pools has rendered
some initial design assumptions \cite{Buterin2015} obsolete today.
Among multiple findings of our study that support this claim,
we observed that the usual 12-block
confirmation rule of Ethereum may not provide the strong probabilistic
guarantees that are promised by probabilistic
analysis that unrealistically rely on a flat and large universe of
individual miners.
More concretely, we observed that the centralization of most mining power on
Ethereum has already enabled alarmingly long sequences of consecutive blocks
(lengths of 8, 9 and even 14 blocks) generated by a single mining pool. This emphasizes that, for permissionless blockchain protocols whose design
allows mining pool-like extensions, these need to always be considered
as first-class elements of the ecosystem at the earliest design stages.
However, we observe that the underlying system model considered by many
research papers that study or propose new blockchain implementations
omits mining pool organizations from
their underlying system models ~\cite{InfPropagaion--02, OnAvailability--01, MeasEthPeers}. This common practice should be avoided by the research community.
\jpb{it would be cool if we supported the above sentence with concrete citations to papers that are examples of this practice. However, probably we don't have the time for that.}

Our study also 
revealed that some key trends changed in a short time span, such as: 
i) the median waiting time for 12 blocks decreased from 200 seconds to 189 seconds~\cite{OnAvailability--01};
ii) a substantial increase in out-of-order committed transactions from 6.18\% to 11.54\%~\cite{OnAvailability--01};
iii) the proportion of forked blocks increased by more than 1\% and fork lengths increased as well.
This confirms that large-scale permissionless blockchain systems 
are eminently dynamic, and highlights the importance of
studies like ours to take place regularly. We make our tools available to encourage this.
Finally, we have systematically confirmed that the geographical location of
a node has a consistent impact on the level of service that node gets from the
system. More than a symptom that the set of nodes is not evenly spread across
the globe, this reflects the fact that a large portion of Ethereum's activity now
 depends on a small and poorly dispersed subset of nodes that comprises the gateways of the major mining pools.
 This stresses the importance of multi-observer measurement approaches when characterizing permissionless blockchains, as followed by our study.

\vspace{-0.2cm}
\section{Conclusion}
\label{conclusions}
We have described our experience 
studying the Ethereum network from several
geographically dispersed observation points.
We identified previously undocumented forms of selfish behavior and showed that the prevalence of powerful mining pools exacerbates the geographical impact on block propagation delays. 
We provide a set of open measurement and processing tools, as well as
the data set of the collected measurements, to promote further research on permissionless blockchains.
%


\IEEEtriggeratref{17}
\bibliographystyle{IEEEtran}
\bibliography{references}

\end{document}